\documentclass[aps,twocolumn,prb]{revtex4}
\makeatletter
\def\@dotsep{4.5}
\makeatother

\usepackage{graphicx}
\usepackage{amssymb}
\usepackage{natbib}
\usepackage{amsmath}
\usepackage{hyperref}

\newcommand{\comment}[1]{}

\begin{document}

\title{Soliton instability and fold formation in laterally compressed few-layer graphene}

\author{Amauri Lib\'erio de Lima$^{1}$}
\author{Lucas A. M. M\"{u}ssnich$^{2}$}
\author{Ta\'{i}se M. Manhabosco$^{1}$}
\author{H\'elio Chacham$^{3}$}
\author{Ronaldo J. C. Batista$^{1}$}
\author{Alan Barros de Oliveira$^{1}$}

\affiliation{$^{1}$ Departamento de F\'{i}sica, Universidade Federal de Ouro Preto, Ouro Preto, MG, 35400-000, Brazil.\\
$^{2}$ Departamento de F\'{i}sica-Matem\'atica, Universidade de S\~ao Paulo, S\~ao Paulo, SP, 05314-970, Brazil.\\
$^{3}$ Departamento de F\'{i}sica, Universidade Federal de Minas Gerais, Belo Horizonte, MG, 30123-970, Brazil.}

\date{\today}
\begin{abstract}
We investigate -- through simulations and analytical calculations -- the consequences of 
uniaxial lateral compression applied to the upper layer of few-layer graphene. The simulations of compressed graphene show that strains larger than 2.8 \% induce soliton-like deformations that further develop into large, mobile folds. Such folds were indeed experimentally observed in graphene and other 
solid lubricants
two-dimensional materials.  Interestingly, in the soliton-fold regime the shear stress decreases with the strain s, initially as $s^{-2/3}$ and rapidly going to zero. Such
instability is consistent with the recently observed negative dynamic compressibility of 
two-dimensional materials. We also predict that the curvatures of the soliton-folds are given by $r_c = \delta\sqrt{\beta/2\alpha},$ where $1\le \delta \le2,$ and $\beta$ and $\alpha$ are respectively related to the layer bending modulus and to the exfoliation energy of the material. This finding might allow experimental estimates of the $\beta/\alpha$ ratio of two-dimensional materials from fold morphology.
\end{abstract}

\maketitle

\section{Introduction} 
\label{intro}

Graphene has gained much attention from the scientific community 
since its discovery because of its unique features. For example, it has been
considered as a possible building block for circuit components due to its particular electronic properties. 
Ideally, deposited graphene should be perfectly flat.
However, although graphene has one of the highest known Young's modulus, it has small bending modulus.\cite{Barboza2009} In real applications, graphene sheets commonly present ripples and folds,
\cite{Parga2008,Diaye2006,Lui2009,Neto2009}
which 
may
change its electronic
structure.
For example, theoretical studies suggest that folded graphene under external magnetic fields  act as interferometer: it suffers the interference due to the interplay between gauge fields created by the fold and the external fields in the region of the fold.\cite{Rainis2011}
Zheng \emph{et al.} have shown that the calculated Young's modulus, tensile strength, and fracture strain of folded graphene are comparable to those of graphene, while the compressive strength and strain are much higher than those of
planar
graphene.\cite{Zheng2011} 

Folds have been observed in several conditions and in a variety 
of forms. They were seen in the top layers of graphite\cite{Hiura1994,Roy1998} whereas edge folds in suspended graphene have been reported as well.\cite{Meyer2007,Warner2009,Huang2009,Liu2009,Zhang2010} 
By concomitantly applying compressive and shear stresses through an atomic force microscopy tip upon few layer graphene, Barboza and collaborators obtained structures which appear to be single- and multi-folded graphene.\cite{Barboza2012}
Multiply folded graphene -- termed grafold by Kim and coworkers\cite{Kim2011} -- were indeed confirmed to exist and, not surprisingly, its novel electronic structure can be quite different in comparison to the flat graphene.\cite{Kim2011,Zhu2012,Xie2012}

Theoretical models suggest that folds in graphene can change its chemical affinity, since curvatures induce deformations in the $\sigma$-bonds of the lattice.  Such out-of-plane deformed bonds 
could transfer charges to $\pi$-orbitals which induce localized dipole moments 
in the graphene surface.\cite{Feng2009} This property could lead
to localized selective functionalization of atoms and molecules. For example, 
Tozzini and collaborators have shown that storage (through adsorption) and release of hydrogen can in principle be obtained by exploiting and controlling the corrugation of individual layers of graphene.\cite{Tozzini2011}
Storage of molecules can also be achieved by  wrapping chemical species 
into graphene folds as sandwiches.\cite{Kim2011,Yuk2011,Monthioux2007}

Given the relevance of graphene folds, it is important to understand the physics behind the folding process. 
Bending orientation, defects, and contamination are probably determinant on such a process.
\cite{Qi2010,Bunch2011,Su2011,Pang2012,Catheline2012, Ortolani2012} 
In this work we investigate  -- trough molecular dynamics and theoretical calculations --
laterally compressed graphene bilayers.  At low strains we observed soliton-like structures 
that evolve into mobile folds with increasing strains. Our results include the derivation of curvature 
radii of some of the main structures formed during compression in terms of
exfoliation and bending energies, $\alpha$ and $\beta$, respectively. Our results
can be applied to any solid lubricant, such as 
molybdenum disulfide
and 
hexagonal boron nitride
. 

This work goes as follows.  In Section \ref{MD} we describe the molecular dynamics methodology and the main fold structures that result from the simulations. In Section \ref{model} we develop analytical models for the fold structures. 
Section \ref{simresults} is destined to discuss 
the simulation results and comparisons with the analytical models. In Section \ref{conc} we present our conclusions. 


\section{Molecular dynamics: methodology and fold structures} 
\label{MD}

Molecular dynamics techniques 
were used as implemented in the package LAMMPS.\cite{lammps}
Carbon atoms were modeled classically using the adaptive intermolecular reactive empirical bond order (AIREBO) potential  for the C-C interaction.\cite{AIREBO} Our system is composed of two graphene layers, each one containing 1600 atoms. 
The bottom layer was kept ``frozen'' during all simulations, i.e., 
the resultant  force on every atom of such layer was set to zero. Both ends of the top layer were
also maintained frozen: the resultant 
forces acting upon 32 atoms of each extrema 
were kept zero in all simulations. Periodic boundary condition was used in 
the $y$ direction, while directions $x$ and $z$ were finite. 
The dimensions of the layers were 207.0 and 18.1  \AA~ in
the $x$ and $y$ directions, respectively. The equilibrium distance between layers 
was found to be around 3.4 \AA.

Simulations were performed in the canonical ensemble.
The Nos\'e-Hoover thermostat\cite{Nose1985,Hoover1985} 
as implemented by Shinoda and collaborators\cite{Shinoda} was used in order to keep 
the temperature $T=10$ K. The timestep used was 0.001 ps. 

\begin{figure}
 \centering
 \includegraphics[clip=true,scale=0.28]{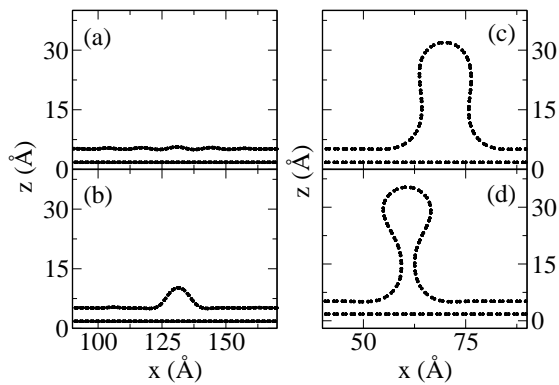}
 \caption{Projection of the graphene bilayer into the $xz$ plane obtained through simulations. (a) The
 moment immediately before the soliton formation at strain $s=2.8$ \% (see Sec. \ref{simresults} for
 the definition of the strain). (b) The
 soliton structure and (c) the structure we termed as \emph{standing fold},
 which appear at $s=27.5$ \%. (d) The structure which appear at approximately $s=28$ \%.
 We named it as  \emph{standing collapsed fold}.}
\label{projall}
 \end{figure}

 Compressive strain in the upper layer was 
imposed along the $x$ direction by moving one of its frozen 
edges 
towards the 
another edge at constant velocity  $v=dx/dt = 0.1$ \AA/ps in the 
$x$ direction.
By increasing the strain, different 
structures are formed in the upper layer. Here
we focus on those shown in Fig. \ref{projall}.
In Fig. \ref{projall}(a)
it is possible to see the $xz$ projection of the upper layer in the moment 
immediately before 
a soliton-like structure appears. The soliton is shown in the Fig. \ref{projall}(b). 
By further increasing
$s$,
two distinct structures appear in sequence. The first is
the \emph{standing fold}, shown in Fig. \ref{projall}(c).
The second is
the \emph{standing collapsed fold}, as
seen in Fig. \ref{projall}(d). For studying such 
structures, we developed theoretical models which are detailed 
in 
the next section. 

\section{Analytical model} 
\label{model}

Our model consists of a continuum 2D material, ideally deposited on a substrate that is parallel to the $xy$ plane. Wrinkles may appear parallel to the $y$ direction, such that the local height $z$ is a function of $x$ only. The 2D material is incompressible but can be bent, with a bending modulus $\beta$ defined such that the curvature energy per unit length, for a given curvature radius $r$, is given by $e_\mathrm{C}=\beta/r^{2}$. We also consider that the binding energy per unit area between the 2D material and the substrate is given by $e_\mathrm{S}=\alpha$. 

\subsection{The soliton structure}

\begin{figure}
 \centering
 \includegraphics[clip=true,scale=0.3]{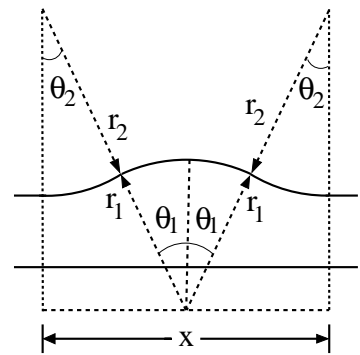}
 \caption{Model for the soliton formation. 
$r_1$ and $r_2$ correspond to the
 top and basis radii, respectively, along with its corresponding angles, $\theta_1$ and $\theta_2$.}
\label{modelo-fig1}
\end{figure}

Let us first consider the soliton-like structure shown in Fig. \ref{projall}(b). We model such structure with three circle segments as shown in Fig. \ref{modelo-fig1}. Considering the continuum model described above, the formation energy per unit length (along $y$) of such a soliton is given by

\begin{align}
\epsilon&=  2\alpha(r_1\theta_1+r_2\theta_2)+2\beta\left(\frac{\theta_1}{r_1}+\frac{\theta_2}{r_2}\right).
\label{eq:energyrope}
\end{align}

As a result of the soliton formation, the 2D material will have an apparent reduction in length, along $x$, of magnitude $\mu$. From Fig. \ref{modelo-fig1}, $\mu$ is given by 

\begin{align}
\mu &= 2(r_1\theta_1+r_2\theta_2)-2\left(r_1\sin\theta_1+r_2\sin\theta_2\right),
        \label{eq:ropemass}
\end{align}

If we consider small angles $\theta_1$ 
and $\theta_2$, we can approximate Eq. (\ref{eq:ropemass}) as

\begin{equation}
\mu \approx \frac{r_1\theta_1^3}{3} + \frac{r_2\theta_2^3}{3}.
\label{eq:ropemasssimple}
\end{equation}

Defining variables $q_i\equiv2\beta\theta_i/r_i$ and $t_i\equiv2\alpha r_i\theta_i$, with $i=1,2$, Eqs. (\ref{eq:energyrope}) and (\ref{eq:ropemasssimple}) can be rewritten as

\begin{equation}
\epsilon=t_1+t_2+q_1+q_2
\label{eq:energyropesimple}
\end{equation}

\noindent and

\begin{equation}
\mu= \frac{1}{24\alpha^2 \beta}\left(q_1 t_1^2+q_2 t_2^2\right).
\label{eq:ropemodified}
\end{equation}

The profile of the soliton can be found by minimizing its energy, Eq. (\ref{eq:energyropesimple}), with $\mu=$ constant. We find

\begin{align}
t_1=t_2=\left(24\alpha^2 \beta \mu \right)^{1/3} & \mathrm{and}& q_1=q_2=\frac{1}{2}t_1.
\label{eq:tisi}
\end{align}

Equations (\ref{eq:tisi})  lead to

\begin{align}
&r_1=r_2=\sqrt{2}\sqrt{\frac{\beta}{\alpha}} \label{eq:r1r2}\\
&\theta_1 (\mu) = \theta_2 (\mu)= \left( \frac{3}{2}\sqrt{\frac{\alpha}{2\beta}}\right)^{1/3}\mu^{1/3}  \label{eq:theta1theta2}\\
&\epsilon(\mu) = 3\left(24 \alpha^2 \beta \right)^{1/3}\mu^{1/3} \label{eq:energyfinal}\\
&l(\mu) =  4 \left(\frac{3\beta}{\alpha}\right)^{1/3}\mu^{1/3}\label{eq:l}.
\end{align}

From  Eqs. (\ref{eq:r1r2})-(\ref{eq:l}) we see that the 
soliton radii $r_1$ and $r_2$ are independent of $\mu$ 
(thus must be strain independent), whereas  the angles $\theta_1$ and $\theta_2$
scale with $\mu^{1/3}$. This interesting behavior suggests that for $\mu \rightarrow 0$
the soliton localizes and disappears without ``flattening''. Another interesting behavior is that of the tension $f$ (force per unit length) necessary to maintain the soliton at a given $\mu$. From Eq. (\ref{eq:energyfinal}), we  obtain 

\begin{equation}
f(\mu)=-d\epsilon/d\mu=K \mu^{-2/3},
\label{eq:f}
\end{equation}

\noindent with $K=-\left(24 \alpha^2 \beta \right)^{1/3}$. That is, the magnitude of $f$ reduces with increasing soliton size, and it tends to infinity as the soliton disappears.

\subsection{The standing fold structure}

\begin{figure}
 \centering
 \includegraphics[clip=true,scale=0.25]{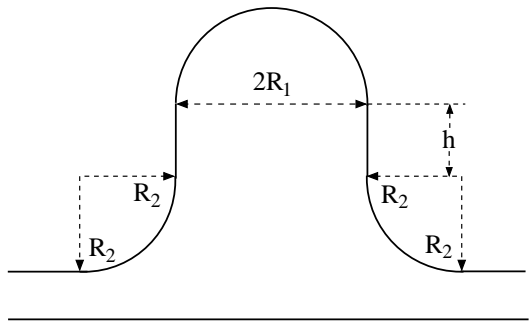}
 \caption{Model for the \emph{standing fold} structure. It has
 a top part modeled as a semi-circle with radius $R_1$ connected to the basis (quarter circles with radius $R_2$) by straight lines with length $h$.}
\label{modelo-fig3}
\end{figure}

For larger values of compressive strain, the soliton structure evolves to the pattern shown in Fig. \ref{projall}(c),  and
schematically shown in Fig. \ref{modelo-fig3}. We model
this structure as follows. The profile of the top part is composed by a semi circle with radius  $R_1$.
A stem is formed by two straight lines with length $h$, and the basis is formed by quarter circles with radii $R_2$. 
The formation energy of such a structure is given by

\begin{equation}
E=\left[\pi \left(R_1+R_2\right) + 2h \right]\alpha + \beta \pi \left(\frac{1}{R_1}+\frac{1}{R_2} \right).
\label{energyfold}
\end{equation}

The  net length to form this structure can be written as 

\begin{equation}
\mu = (\pi-2)(R_1+R_2)+2 h. 
\label{massfold}
 \end{equation}

Equations (\ref{energyfold}) and (\ref{massfold}) reduce to

\begin{equation}
E = \alpha \left[ \mu + 2\left(R_1+R\right) \right] + \beta \pi \left(\frac{1}{R_1}+\frac{1}{R_2} \right).
\label{energyfold2}
 \end{equation}

After minimizing Eq. (\ref{energyfold2}) with respect to $R_1$ and $R_2$ with $\mu$ constant one obtains

\begin{equation}
R_1 = R_2 = \sqrt{\frac{\pi}{2}} \sqrt{\frac{\beta}{\alpha}}.
\label{eq:radius}
 \end{equation}

\subsection{The standing collapsed fold structure}

\begin{figure}
 \centering
 \includegraphics[clip=true,scale=0.28]{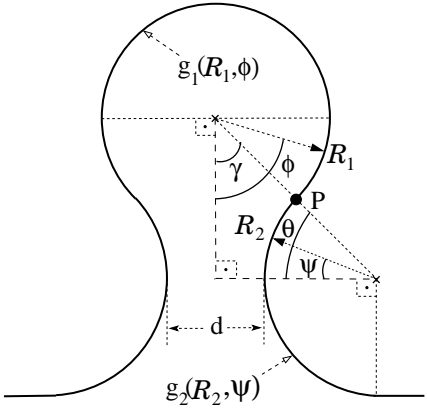}
 \caption{The proposed model for the \emph{standing collapsed fold}. The top and the basis are arcs of circle with radii $R_1$ and $R_2$, respectively.}
\label{modelo-fig5}
\end{figure}

By further increasing the strain, the next type of structure observed for compressed graphene is schematically shown in Fig. \ref{modelo-fig5}. It corresponds to the structure shown in Fig. \ref{projall}(d) obtained through simulations. We  call this structure as \emph{standing collapsed fold}. 
We modeled such an structure as an arc of circle of radius ${\cal{R}}_1$, forming
the top of the structure, which is connected to the basis by arcs of circle of radius ${\cal{R}}_2$.
$g_1({\cal{R}}_1,\phi) = {\cal{R}}_1 \phi$ and $g_2({\cal{R}}_2,\psi) = {\cal{R}}_2 \psi$ are the functions which define the head and the basis curves in polar coordinates, respectively. $g_1$ and $g_2$ intercept each other at 
the point P. In this sense, $g_1({\cal{R}}_1,\phi=\gamma)=g_2({\cal{R}}_2,\psi=\theta)$. Since 
$\gamma=\pi/2-\theta$, we find that

\begin{equation}
\theta = \frac{{\cal{R}}_1 \pi}{2 \left({\cal{R}}_1 + {\cal{R}}_2 \right)}. 
\label{eq:theta}
\end{equation}

 The  formation energy is  given by

\begin{equation}
 {\cal E}=\left(\pi + 2\theta \right)\left[\left({\cal{R}}_1 + {\cal{R}}_2\right)\alpha+\left(\frac{1}{{\cal{R}}_1}+\frac{1}{{\cal{R}}_2}\right)\beta\right].
 \label{eq:energyronaldo}
 \end{equation}

The variable $\theta$ can be eliminated from Eq. (\ref{eq:energyronaldo}) 
with help of Eq. (\ref{eq:theta}). Thus the energy reduces to

\begin{equation}
{\cal E}=\alpha\pi(2{\cal{R}}_1+{\cal{R}}_2)+\beta\pi\left(\frac{1}{{\cal{R}}_1}+\frac{2}{{\cal{R}}_2}\right).
\label{eq:encofold}
\end{equation}

By minimizing with respect to ${\cal{R}}_1$ and ${\cal{R}}_2$, i.e., performing $\partial{\cal E}/\partial R_{1}=0$ and $\partial{\cal E}/\partial R_{2}=0,$ one finds

\begin{align}
{\cal{R}}_1=\frac{\sqrt{2}}{2}\sqrt{\frac{\beta}{\alpha}}  &&\mathrm{and}&& {\cal{R}}_2=\sqrt{2}\sqrt{\frac{\beta}{\alpha}}.
 \label{eq:R1}
\end{align}

It is possible to calculate the length of the rope, ${\cal L}$, detached from the substrate 
as a function of $\alpha$ and $\beta$. By simple inspection of Fig. \ref{modelo-fig5}
we see that 

 \begin{equation}
 {\cal L}=\left(\pi + 2\theta \right)({\cal{R}}_1 + {\cal{R}}_2).
 \label{eq:length}
 \end{equation}

Once ${\cal{R}}_2=2{\cal{R}}_1$,  Eq. (\ref{eq:theta}) gives $\theta=\pi/6$.
This result, along with Eqs. (\ref{eq:R1}) and (\ref{eq:length}) , gives

\begin{equation}
 {\cal L}=2 \sqrt2 \pi \sqrt{\frac{\beta}{\alpha}}.
 \label{eq:length3}
 \end{equation}

Another important result we can derive from this model is the minimum distance between
the base arcs $d$. From Fig. \ref{modelo-fig5}, we find that

\begin{align}
d =2\left[\left({\cal{R}}_1+{\cal{R}}_2\right)\cos\theta-{\cal{R}}_2\right]= \nonumber\\ 
     \frac{\sqrt{2}}{2}\left(3\sqrt{3} - 4 \right)\sqrt{\frac{\beta}{\alpha}},
\label{dneck}
\end{align}

\noindent where we have used Eq. (\ref{eq:R1}) and  $\theta=\pi/6$.

\section{Simulation results} 
\label{simresults}

We define the compressive strain in the upper layer as 
$s(t)=v t / L_x$, where $L_x$ is the dimension of the layer
in the $x$ direction and $t$ is the time.
In this sense, the strain is zero at the initial time, $t=0$,
and it is maximum ($s=100$ \%) when $vt=L_x$.

In order to characterize the structures shown in Fig. \ref{projall}, namely, the
soliton, the \emph{standing fold} and the \emph{standing collapsed fold}, we calculated the tension
 versus the compressive strain in the upper layer. 
 The tension was calculated as $P_{xx} L_y$, where
 $P_{xx}$ is the virial contribution for the component of the stress tensor in the $x$ direction,
 and $L_y$ is the dimension of the simulation box in the $y$ direction.
 $P_{xx}$ is given by  
 
 \begin{equation}
 P_{xx} = \frac{1}{V}\sum_{i=1}^N x^i f^{i}_x
 \end{equation}
 
 \noindent where $V$ is the volume of
 the simulation box, $N$ is the number of particles, $x^i$ is
 the coordinate $x$ of particle $i$, and $f_x$ is the
 component of the resultant force acting on particle $i$ in the $x$ direction.
 
\begin{figure}
 \centering
 \includegraphics[clip=true,scale=0.3]{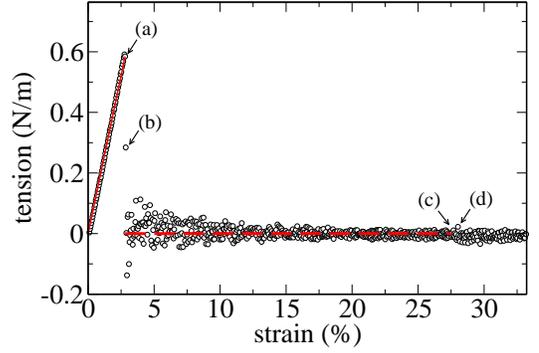}
 \caption{Tension in the upper graphene layer as a function of the induced strain $s$.
 At $s=0$ the tensile force upon the graphene layer is zero. By increasing $s$ 
 the graphene behaves as an elastic medium until the soliton formation at $s=2.8$ \%.
 At this point, internal forces are released which causes the discontinuity seen in the figure.
 For $s>2.8$ the curve decays as $s^{-2/3}$ which is explained by our theoretical model (see Sec. \ref{model}). Arrows (a)-(d) correspond to the instants where the structures seen in Fig. \ref{projall} appear.}
\label{history}
 \end{figure}

 Our tension vs. strain results are summarized in 
  the Fig. \ref{history}. Arrows (a)-(d) indicate the
 instants where the structures shown in Fig. \ref{projall} appear. In the first stages of compression, the layer behaves 
 elastically, with linear response to the applied strain. 
 The straight, continuous line corresponds to a linear fitting
 through the data whose angular coefficient was
 found to be 0.21 N/m.
 After reaching a maximum tension $\tau$ value, at $s=2.8$ \%, 
 the soliton appears, releasing internal forces, which explains the discontinuity in the stress-strain curve. 
 Further increasing $s,$ the tension decays towards zero 
 with a $s^{-2/3}$ dependence. Such a dependence is explained 
 by our analytical model, Eq. (\ref{eq:f}).
 The fitting for $s>2.8$  \% (dashed line) was made by using a function 
 in the form $\tau=\kappa s^{-2/3}$, where $\kappa = 0.003$ N/m.
 
\begin{figure}
 \centering
 \includegraphics[clip=true,scale=0.28]{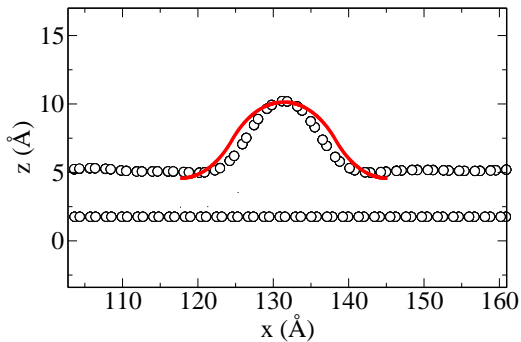}
 \caption{Dots: the soliton as obtained in our simulations. Continuous curve: result from the soliton model (Eqs. (\ref{eq:r1r2}) and (\ref{eq:theta1theta2})).}
\label{simsoliton}
 \end{figure}

\begin{figure}
 \centering
 \includegraphics[clip=true,scale=0.28]{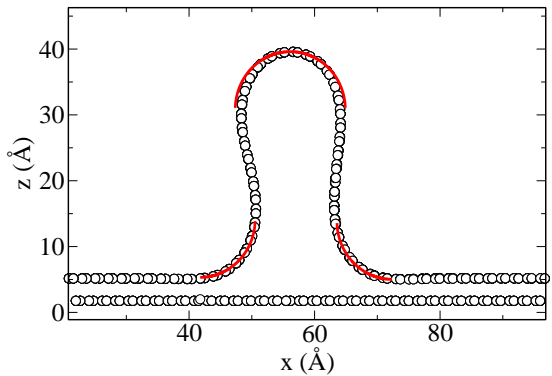}
 \caption{Dots: the \emph{standing fold} structure, as obtained in the simulations (also seen in Fig. \ref{projall}(c)). Continuous curves: curvature radii as obtained from the analytical model, see Eq. (\ref{eq:radius}) and relation (\ref{alphabeta}).}
\label{simfold}
 \end{figure}

\begin{figure}
 \centering
 \includegraphics[clip=true,scale=0.28]{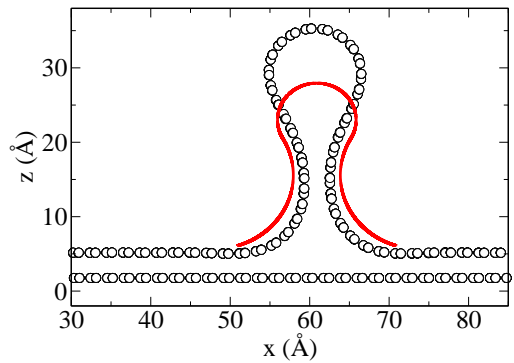}
 \caption{Dots: \emph{collapsed standing fold} structure, also seen in Fig. \ref{projall}(d). Countinuous curve: analytical model, see Eqs. (\ref{eq:R1}) and (\ref{alphabeta}).}
\label{simcolfold}
 \end{figure}

We have also compared  predictions of our theoretical models with
the results from the simulations. Figures \ref{simsoliton}, \ref{simfold}, and \ref{simcolfold}
show the structures seen in Figs. \ref{projall}(b),  \ref{projall}(c), and  \ref{projall}(d), 
respectively, superimposed with continuous curves that
correspond to curvatures as obtained by our models.  

The results from the models depend on the binding energy 
and the bending modulus of the graphene ($\alpha$ and $\beta$, respectively), always
in the form $\sqrt{\beta/\alpha}$, which turned to be an \emph{intrinsic length scale}. 
There have been several attempts to determine  the graphite binding energy, both experimentally
\cite{Girifalco1956,Benedict1998,Zacharia2004} and theoretically.\cite{Charlier1994,Trickey1992,Rydberg2003,DiVincenzo1983,Schabel1992}
To the best of our knowledge, the most recent, direct graphite binding energy  measurement is given by Liu \emph{et al.}, who have obtained $\alpha=31$ meV/atom.\cite{Liu2012} For the graphene bending modulus,
the most direct measurement is due to Barboza and collaborators \cite{Barboza2009}  who have
 found $\beta=1.64$ eV\AA$^2$/atom. Thus,  the \emph{intrinsic length scale}  seen in our models is given by
 
 \begin{equation}
 \sqrt{\beta/\alpha} = 7.0~\mathrm{\AA}.
  \label{alphabeta}
 \end{equation}

The soliton model (Fig. \ref{modelo-fig1}) predicts 
that the radii of both basis and top must be the same. We have found
such curvature radii $r_1 = r_2=$ 9.9 \AA~[see Eq. (\ref{eq:r1r2})]. In order to compare the model and simulation results,
we have used Eq. (\ref{eq:theta1theta2}) with $\mu=3.0$ \AA~(estimated from simulations) for drawing 
the soliton as predicted by our model. This leads to $\theta=44^{\circ}$. The resulting curve is seen in Fig. \ref{simsoliton}
as a line, while the circles mark the position of carbon atoms, obtained by simulations. 

Figure \ref{simfold} shows the \emph{standing fold} structure as obtained by simulations. This
structure is seen when the strain $s$ is around 27.5 \% (see Fig. \ref{history}). 
We have modeled such
an structure as having circumference arcs in the top (with radius $R_1$) and in the basis (radius $R_2$)  (see Fig. \ref{modelo-fig3}). Top and basis are connected by straight lines with length $h$. We obtained 
$R_1=R_2=$ 8.8 \AA~[see Eq. (\ref{eq:radius})]. Figure  \ref{simfold}
shows the model predictions (lines) along with 
simulations results (circles). We see that in spite the simplicity of the model
it gives reasonable results compared to simulations without the need of any information from it
other than the bending modulus and the binding energy values. 
Therefore, such models can be used to make predictions on the fold geometry of other solid lubricants
 and vice-versa, that is, to predict the value of the ratio $\beta/\alpha$ based purely on fold geometry. 

 Finally, Fig. \ref{simcolfold}
shows the \emph{standing collapsed fold}, which was modeled as having 
$R_1$ for the top radius and $R_2$ for the basis (as shown in Fig. \ref{modelo-fig5}). We concluded 
 that the condition which minimizes the energy of the 
collapsed fold is $R_2 = 2 R_1$, given by Eqs. (\ref{eq:R1}). Our findings are 
$R_1 = 4.95$ \AA~and $R_2=9.9$ \AA.
Note that the laterals of this structure tend to approach the 
bilayer distance (around 3.4 \AA) for big strains as expected.
The result from the model for such a distance is $d =$5.9 \AA~ [see Eq. (\ref{dneck})], 
which lies in the same order of magnitude. Since we have
not considered van der Waals interaction between layers, it is not surprising
we have found a bigger value for such a distance than the expected value of $\sim$3.4 \AA.

It is worth to mention we  have found from the \emph{standing collapsed fold} model
 an expression for the length of the
layer which is detached from the substrate as a function of $\alpha$ 
and $\beta$, in the moment it is formed [Eq. (\ref{eq:length3})]. 
We estimate such a value from simulation as 84 \AA,
while Eq. (\ref{eq:length3}) gives 62 \AA.

\begin{figure}
 \centering
 \includegraphics[clip=true,scale=0.24]{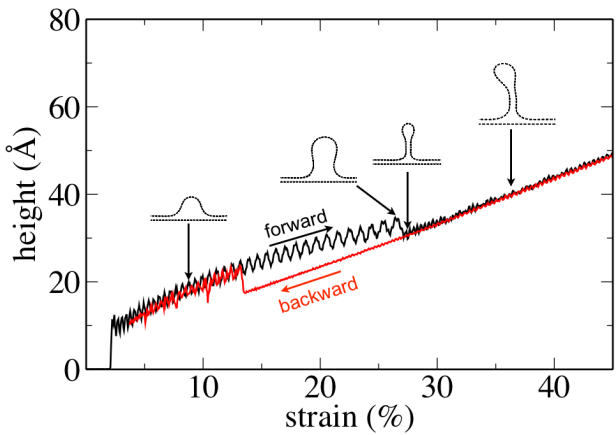}
 \caption{(color online) Height of the fold structures, relative to the unstrained upper layer, versus the induced strain. 
 The black curve refers to simulations with increasing strain, and red curve refers to simulations with decreasing strain. The \emph{standing collapsed fold}
 is formed at $s=$ 28 \% in the increasing strain simulation, but it persists for strains up to $s=$ 13.4 \% 
 in the decreasing strain simulation, evidencing a hysteresis effect.}
\label{histerese}
 \end{figure}

In order to investigate the stability of the \emph{standing collapsed fold} 
 structure, we proceeded as follows. As stated in Sec. \ref{MD},
strain was induced  in the $x$ direction of the upper layer by moving one of its extremities towards the opposite one at
constant velocity. After a certain maximum strain, we inverted the movement direction, keeping the 
velocity modulus, which continuously reduces the strain. During this ``forward-backwards" process, we
monitor the height of the structures in relation to the upper layer against the induced strain.  The results are summarized in Fig. \ref{histerese}. From this figure, we see all the stages approached in this work, namely, the soliton
(see the jump at around $s=$ 2.8 \% which characterizes its appearance) and its continuation until the \emph{standing fold} takes place  at around $s=$27.5 \%. 
At $s=28$ \% we observe the transition from  \emph{standing fold} to \emph{standing collapsed fold} with linear
dependence between height and strain for $s>$ 28 \%. When the direction of the movement is
inverted, the \emph{standing collapsed fold} becomes stable for strains below $s=28$ \%. Indeed,
the ``uncollapsing'' transition occurs at $s=$ 13.4 \%. Figure \ref{histerese} has the characteristics of a hysteresis curve,
in which the state of the system depends not only to the strain at a certain time but also to its history.

\section{Conclusions}\label{conc}

This work is an investigation -- through molecular dynamics and analytical calculations -- of a laterally compressed graphene monolayer atop uncompressed graphene (simulating an uncompressed graphite surface).
Under compression, several structures
appear in the top graphene layer. Three structures can be clearly identified: the soliton, the \emph{standing fold}, and the 
\emph{standing collapsed fold} structures. We propose models for each of such structures, and we have determined curvature radii for those structures in terms of $\alpha$ and $\beta$, the exfoliation and bending energies, respectively. Our models indicate that all structures have characteristic radii in terms of $\sqrt{\beta/2 \alpha}$, as seen in Eqs. (\ref{eq:r1r2}),  (\ref{eq:radius}), and (\ref{eq:R1}). This result is general and can be applied to other solid lubricants, like MoS$_2$, talc, and hexagonal boron nitride, for example, to estimate the ratio $\beta/\alpha$ from fold morphology. We have also found that the \emph{standing collapsed fold} shows bi-stability in relation to the strain $s$,
depending on the path for achieving critical strains. Upon increasing strain, such structure appears at $s=$ 28 \%. Once it appears, if the strain is decreased,
the  \emph{standing collapsed fold} remains stable until $s=$ 13.4 \%, showing a hysteresis behaviour. 
\section*{ACKNOWLEDGMENTS}

We thank for financial support from the Brazilian science agencies CNPq, CAPES and FAPEMIG. This work is also partially supported by the project INCT-Nanomateriais de Carbono.

\bibliographystyle{aip}
\bibliography{Biblioteca}

\end{document}